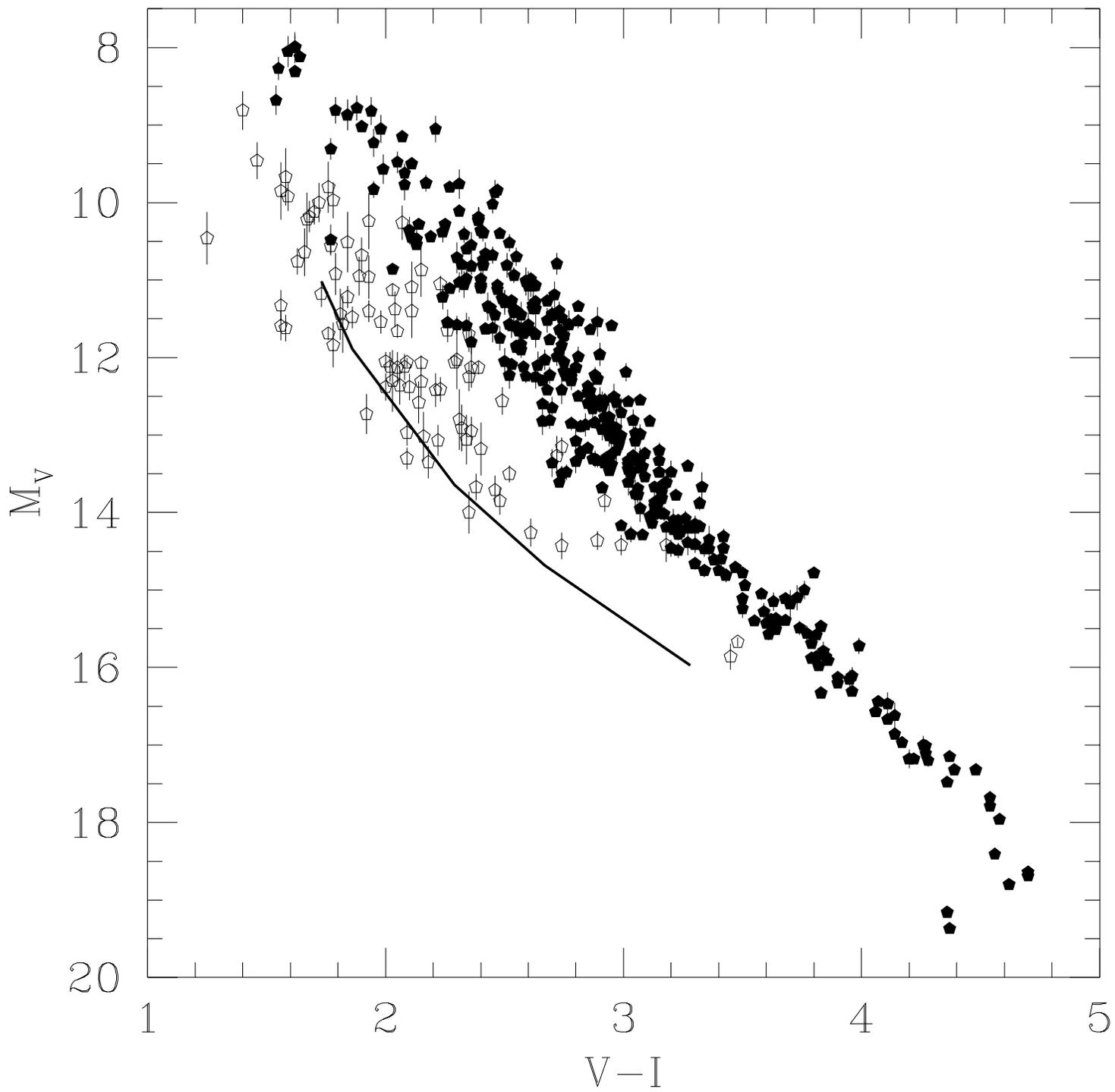

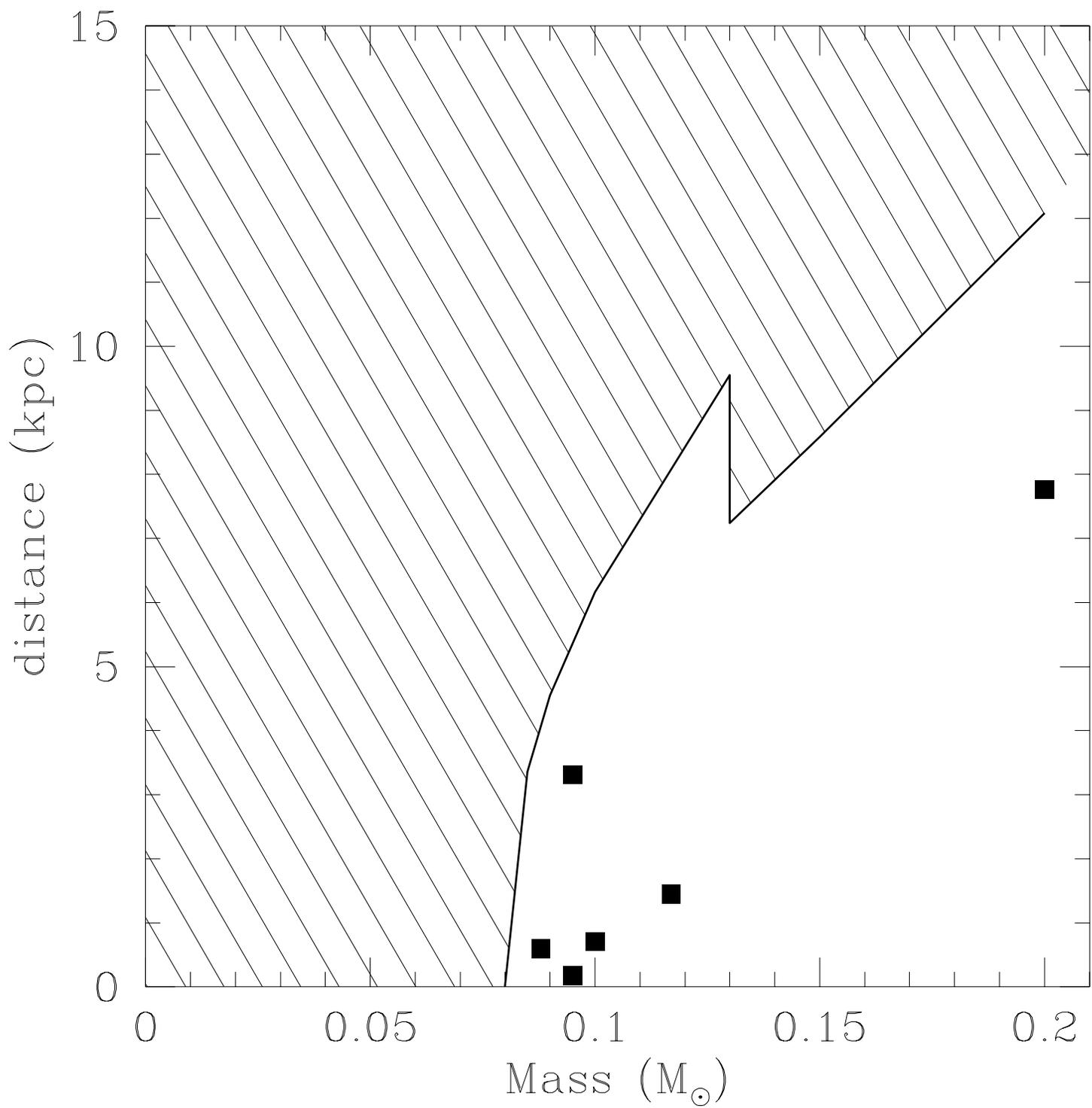

# Analysis of a Space Telescope Search for Red Dwarfs: Limits on Baryonic Matter in the Galactic Halo

**David S. Graff**[1] and **Katherine Freese**[1,2]

1) *University of Michigan, Department of Physics*
*Ann Arbor, MI 48109-1120*
2) *Institute for Theoretical Physics, University of California*
*Santa Barbara, CA 93106*

## Abstract

We re-examine a deep *Hubble Space Telescope* pencil-beam search for red dwarfs, stars just massive enough to burn Hydrogen. The authors of this search (Bahcall, Flynn, Gould & Kirhakos 1994) found that red dwarfs make up less than 6% of the galactic halo. First, we extrapolate this result to include brown dwarfs, stars not quite massive enough to burn hydrogen; we assume a $1/\mathcal{M}$ mass function. Then the total mass of red dwarfs and brown dwarfs is $\leq 18\%$ of the halo. This result is consistent with microlensing results assuming a popular halo model.

However, using new stellar models and parallax observations of low mass, low metallicity stars, we obtain much tighter bounds on low mass stars. We find the halo red dwarf density to be $< 1\%$ of the halo, while our best estimate of this value is 0.14-0.37%. Thus our estimate of the halo mass density of red dwarfs drops to 16-40 times less than the reported result of Bahcall et al (1994). For a $1/\mathcal{M}$ mass function, this suggests a total density of red dwarfs and brown dwarfs of $\sim 0.25$-$0.67\%$ of the halo, *i.e.*, $(0.9 - 2.5) \times 10^9 \mathcal{M}_\odot$ out to 50 kpc. Such a low result would conflict with microlensing estimates by the MACHO group (Alcock *et al.* 1995a,b). We suggest that either the halo mass function must rise very steeply below the hydrogen burning limit, or that the microlensing results should be reinterpreted with a different halo model or mass function.

*Subject Headings: dark matter – Galaxy: stellar content – Galaxy: halo – stars: low mass, brown dwarfs – stars: Population II*

*e-mail:* `graff@umich.edu, freese@mich1.physics.lsa.umich.edu`

**Introduction:** There has recently been a renaissance of interest in halo stars. Halo brown dwarfs (stars just below the hydrogen burning mass) are serious dark matter candidates (Ashman 1992). Gravitational microlensing searches have found evidence of MACHO's, discrete masses in our Galaxy (Alcock *et al.* 1993, 1995a,b; Aubourg *et al.* 1993; Udalski *et al.* 1993). These objects are likely to be brown dwarfs and red dwarfs (stars with mass just above the hydrogen burning mass limit). Assuming a spherical King model halo, the MACHO group calculated that the Milky Way halo might be composed of $18.8^{+38.4}_{-14.8}\%$ (95% c.l.) lensing objects (Alcock *et al.* 1995a). However, with different halo models, one can interpret their results to imply a halo ranging from no lensing objects to a halo composed entirely of lensing objects, although 100% MACHO halo models are constrained (Gyuk, Gates & Turner 1995, Alcock *et al.* 1995b).

Mature brown dwarfs are too dim to be observed directly by telescope. But brown dwarfs and red dwarfs may be formed by the same physical process. Thus, by examining the mass function of red dwarfs, it may be possible to extrapolate below the hydrogen burning limit and study the contribution of brown dwarfs to the galactic halo.

Recently, Bahcall, Flynn, Gould, and Kirhakos (1994, henceforth BFGK) published the results of a Hubble Space Telescope search for red dwarfs. They placed a conservative upper bound of 6% on the mass of the halo in stars with $(V - I)$ colors between two and three. This result has been widely used to defeat the idea of red dwarfs as a dominant component of the halo. However, as we shall see, Pop. II stars with these colors lie in the narrow mass range between $0.087 \mathcal{M}_\odot < \mathcal{M} < 0.13 \mathcal{M}_\odot$. It is possible that the total mass of red dwarfs and brown dwarfs over their much larger possible mass range may still be significant.

We can estimate the total mass of brown dwarfs if we assume a mass function $\frac{dN}{d\mathcal{M}}$. We take a power law with a possible cutoff, $\frac{dN}{d\mathcal{M}} \propto \mathcal{M}^{-\alpha}$. The total mass in a mass range $\Delta$ is $M_{\text{tot}} = \int_\Delta \mathcal{M} \frac{dN}{d\mathcal{M}} d\mathcal{M}$. For stars with mass $\sim \mathcal{M}_\odot$, $\alpha$ is 2.35 as found by Salpeter (1955). For red dwarfs, $\alpha$ is much harder to measure. In this paper, we assume $\alpha = 1$. While Richer and Fahlman (1992) measured a steeply rising halo luminosity function implying a larger value of $\alpha$, Dahn *et al.* (1995) report a luminosity function that rises then falls with decreasing mass, implying $\alpha \sim 1$ (Graff *et al.*, in progress). In addition, $\alpha \sim 1$ is consistent with measurements of the disk IMF by Kroupa (1995) and Henry (1995) and the IMF of globular cluster NGC 1261 by Zoccali *et al.* (1995). For $\alpha = 1$, the total mass in a particular mass range is proportional to that mass range, $M_{\text{tot}} \propto \Delta$. Thus, when we extend the BFGK measurement to cover the entire mass range $0 < \mathcal{M} < 0.13 \mathcal{M}_\odot$, we derive a halo MACHO fraction of $\leq 16\%$, comfortably consistent with the claim by the MACHO group of $18.8^{+38.4}_{-14.8}\%$ (95% accuracy) for the halo MACHO fraction (Alcock *et al.* 1995a).

However, we shall see that BFGK were extremely cautious in vinterpreting their data. They assumed that all red dwarfs were as dim as the dimmest seen *high metallicity* star. By assuming that the stars seen belong to a bluer, brighter population of low metallicity stars, and making less conservative, but still reasonable assumptions, we will derive a much lower estimate of 0.14-0.37% of the halo. This low result will make it difficult to tie the relatively high halo MACHO fraction reported by the MACHO group with a slowly rising mass function. We will suggest that either the assumptions made by the MACHO group about the halo structure and mass function are wrong, or else the mass function must begin to rise steeply below the hydrogen burning limit.

**Deep Imaging Searches for Halo Red Dwarfs:** Distinguishing between dim stars and galaxies by morphology requires the precision of the Hubble Space Telescope. BFGK (1994) used the HST to examine a small deep field and measured the relative magnitudes of stars in the $V$ and $I$ bands. The color of a star is reported as the difference in magnitude between these two bands, $(V - I)$. High values of $(V - I)$ correspond to red stars. BFGK found no stars with $(V - I) > 3$ and only 5 stars with $2 < (V - I) < 3$. From this data, they placed an upper limit of the mass fraction of the halo in red dwarfs of 6%.

Boeshaar, Tyson, and Bernstein (1994) made a ground based search for halo red dwarfs. Lacking the Space Telescope's resolving capabilities, they distinguished stars from galaxies by observing in several bands. They found only 4-6 red dwarfs in their field, representing 0.3% of the halo, one twentieth of the BFGK upper limit. We feel that a careful reinterpretation of the important measurement of BFGK is in order. To place limits on the density of stars in the halo, we must know the distance to the observed stars, which in turn requires that we also know the absolute magnitude of stars of that color. This can be found using parallax surveys.

**Parallax Surveys:** The USNO's parallax survey, as initially reported by Monet *et al.*

(1992), has been greatly expanded (Dahn, Liebert, Harris, and Guetter 1995). In Fig. 1, we reproduce the relevant results. There are two sequences of stars in this graph: 1) To the right (red) is the disk main sequence. 2) Just below the disk stars are several "extreme subdwarfs" shown in open circles. These stars all have high proper motions and those that have been observed spectroscopically have extremely low metallicities, $Z \approx .01 Z_\odot$, i.e., $[M/H] \approx -2$. We use the notation $[M/H] \equiv \log(Z/Z_\odot)$. Thus, these extreme subdwarfs are likely to be pop. II halo stars.

**Properties of Low Metallicity Red Dwarfs:** A separate stellar population, to the left (blue) of the high metallicity main sequence in an H-R diagram, is predicted by theoretical studies of low metallicity red dwarfs. Zero metallicity brown dwarfs were studied theoretically by Burrows *et al.* (1993) and Saumon *et al.* (1994). They found that zero metallicity stars are substantially bluer and more luminous than solar metallicity stars of the same mass. High metallicity stars are more massive and more luminous than low metallicity stars of the same color. We will assume that stars in the halo are roughly as old as the galaxy, and are mature; i.e the red dwarfs are on the low metallicity main sequence.

Stars of low, but finite luminosities were modeled by Baraffe, Chabrier, Allard, and Haushildt (1995). They modeled stars with $[M/H] \equiv \log(Z/Z_\odot) = \{0.0, -0.5, -1.5\}$ corresponding to $Z = \{1, 0.3, 0.03\} Z_\odot$. Their models are in agreement with the observations of Monet *et al.* (1992) and Dahn *et al.* (1995), in the sense that the two sets of high metallicity models agree with the observed disk stars and the $[M/H] = -1.5$ models agree with the population of "extreme subdwarfs" as shown in Fig. 1. We will assume that the stars seen by BFGK are from the same population as the "extreme subdwarfs" and can thus be described by the $[M/H] = -1.5$ ($Z = 0.03 Z_\odot$) models of Baraffe *et al.* (1995). If anything, these models may slightly underestimate the absolute magnitude which will cause us to slightly overestimate the contribution to the halo density.

*The Edge of the Main Sequence:* Many authors make the approximation that stellar properties change abruptly for stars with mass at the hydrogen burning limit: i.e., the main sequence consists of stars able to burn hydrogen and has a sharp boundary, or 'edge', as the luminosity drops toward the hydrogen burning limit. A more realistic interpretation is that the properties of stars change asymptotically, instead of abruptly, as the mass approaches the hydrogen burning limit; then the main sequence should not have a well defined edge. However, we will use the term 'edge' to refer generically to the hydrogen burning limit of a sequence of stars of a given metallicity.

In the models of Saumon *et al.*, $Z = 0$ metallicity stars just at the Hydrogen burning limit are slightly heavier and much bluer and $\sim 2$ times more luminous than stars at the edge of the solar metallicity main sequence. This result is in qualitative agreement with the results of the parallax surveys. As can be seen in Fig. 1, the edge of the high metallicity disk main sequence (solid circles) is much redder and dimmer than the edge of the low metallicity halo population (open circles).

The bulk of the halo stars (extreme subdwarfs) in the survey by Dahn *et al.* (1995) are bluer than $(V - I) = 3.0$. There are a few stragglers extending down to $(V - I) = 3.5$ (see Fig. 1). BFGK find stars out to $(V - I) = 2.9$. Baraffe *et al.* (1995) calculate that low metallicity stars can be as red as $(V - I) \sim 3.28$ for $[M/H] = -1.5$. However, the mass-color function becomes very steep near the main sequence edge. In order for a star to have $3.0 < (V - I) < 3.28$, the mass of that star must be $0.085 < \mathcal{M}/\mathcal{M}_\odot < 0.0875$. Since this mass range is so narrow, we should not be surprised that BFGK do not find redder stars. Because the expected number of low metallicity stars with $(V - I) > 3$ is so small, we focus on stars with colors in the range $2 < (V - I) < 3$ in this paper to obtain estimates of their contribution to the mass density of the halo.

**Estimating the Mass Density of Halo Red Dwarfs:** We will assume that all halo stars are of the same population and have the same absolute magnitudes as the extreme subdwarfs seen by the parallax surveys, which are consistent with the low metallicity $[M/H] = -1.5$ red dwarfs calculated by Baraffe *et al.* (1995). These stars have absolute magnitudes $M_I = 10.5$ when $(V - I) = 2$ and $M_I = 12.3$ when $(V - I) = 3$.

The limiting apparent magnitude in BFGK is $I_{\max} = 25.3$ over the color range $(V - I) > 2$. It is $I_{\max} = 24.7$ for $(V - I) < 2$. One can derive the corresponding absolute magnitude $M_I$ for stars of a given color $(V - I)$ by using Fig. 1; thus, $M_I$ is a function of $(V - I)$. From the absolute magnitude and the limiting apparent magnitude, one can derive the maximum luminosity distance $d_{\max}$ as a function of $(V - I)$, $d_{\max} = 10^{[0.2(I_{\max} - M_I) + 1]}$pc. Here $d_{\max}$ ranges from 4.0 kpc for $(V - I) = 3$ to 9.1 kpc for $(V - I) = 2$.

Since BFGK measured a field of 4.4 arcmin$^2$, or $\Omega = 3.723 \times 10^{-7}$ steradians, the number density of stars in the halo with colors between $(V - I) = 2$ and $(V - I) = 3$ is

$$n = \int_{(V-I)=2}^{(V-I)=3} \frac{dN(V - I)}{\frac{1}{3}\Omega d_{\max}^3}. \tag{1}$$

where $dN(V - I)$ is the number of stars detected between $(V - I)$ and $(V - I) + d(V - I)$. The denominator is the volume of the cone searched.

*An Initial Upper Limit on the Mass Density of Halo Red Dwarfs:* We will perform several successive improvements to the bounds on low mass stars obtained from BFGK data; we focus on stars with $2 < (V - I) < 3$. In this section, we follow BFGK in assuming that all detected stars have the luminosity of a star at the very edge of the main sequence. However, whereas they used the stars at the edge of the *high* metallicity main sequence, we use the stars at the edge of the *low* metallicity main sequence for our calculation. Here we obtain an upper limit on the mass density of halo red dwarfs; in the next section we perform further improvements in order to obtain a more precise estimate.

BFGK used the stars at the edge of the *high* metallicity main sequence for their calculation. As discussed previously, these stars are much dimmer ($M_I = 14$) than stars at the edge of the low metallicity main sequence ($M_I = 12.3$). The dimmest low metallicity objects probably more typical of the halo can be seen out to a greater distance than the dimmest high metallicity stars (see Fig. 1). Thus BFGK probably underestimated $d_{max}$ in the denominator of equation (1), thereby overestimated the density of halo stars, and obtained a very conservative bound.

Instead we use the stars at the edge of the *low* metallicity main sequence for our calculation. Then equation (1) is bounded above by

$$n < \frac{3N}{\Omega d_{\max}^3(3)} = 8.4 \times 10^{-5} N \,\mathrm{pc}^{-3}, \tag{2}$$

where $N$ is the total number of stars detected in BFGK between $(V - I) = 2$ and $(V - I) = 3$, and $d_{\max}$ is evaluated at $(V - I) = 3$. Assuming that all these stars have a mass of about $0.1\mathcal{M}_\odot$, this translates into a mass density upper bound of $\rho < 8.4 \times 10^{-6} N \mathcal{M}_\odot \mathrm{pc}^{-3}$.

If we follow BFGK in assuming that the local halo mass density is $\rho_0 \approx 9 \times 10^{-3} \mathcal{M}_\odot/\mathrm{pc}^3$, and if the halo were made entirely of hydrogen burning red dwarfs, we would expect to see about 1100 stars in the sample. Instead, BFGK only found 5 with $(V-I) > 2$. With Poisson statistics this observation corresponds to an upper limit of 10.5 or fewer stars in each search volume at the 95% confidence level (i.e., if the average number in such a field were 10.5, then the probability of seeing $\leq 5$ would be 5%). Thus, red dwarfs at the low metallicity main sequence edge make up less than 0.95% (i.e. $\frac{10.5}{1100}$) of the halo with 95% confidence.

*A More Precise Estimate of the Mass Density of Halo Red Dwarfs:* Our second improvement to the estimate of the red dwarf contribution to the halo is obtained by using properties of individual observed stars rather than by assuming that they all behave like those at the edge of the main sequence. We will also greatly extend the probed mass range by including a sixth star from the BFGK sample, with $(V-I) = 1.7, \mathcal{M} = 0.2\mathcal{M}_\odot$. From the measured color of each star $i$ in the BFGK sample with $1.7 \leq (V-I) \leq 3$, we obtain an absolute magnitude for that star from Fig. 1; then using the apparent magnitude of the star, we can obtain a maximum luminosity distance $d_{\mathrm{max},i}$. Again, we assume that the stars have low metallicities. Each star has an associated cone of volume $\mathcal{V}_i = \frac{1}{3}\Omega d^3_{\mathrm{max},i}$ and a mass $\mathcal{M}_i$ (See Fig. 2). $\mathcal{M}_i$ was derived from the low metallicity $[M/H] = -1.5$ models of Baraffe *et al.* (1995) which are consistent with parallax observations.

We can write the mass fraction of these stars relative to the halo as

$$f = \sum_i \frac{\mathcal{M}_i}{M_{\mathrm{halo},i}}. \tag{3}$$

Here, $M_{\mathrm{halo},i}$ is the total halo mass contained in the observation cone associated with star $i$,

$$M_{\mathrm{halo},i} = \int_{\mathcal{V}_i} \rho_{\mathrm{halo}}(\vec{x}) d\mathcal{V} = \int_0^{d_{\mathrm{max},i}} \Omega r^2 \rho_{\mathrm{halo}}(r, b, l) \, dr. \tag{4}$$

Here $r$ is the distance from the Earth and $b$ and $l$ are galactic coordinates of the observation direction. For BFGK, $b = -51°, l = 241°$. We assume the same model as the standard model of the MACHO group:

$$\rho_{\mathrm{halo}}(R) = \rho_0 \frac{R_0^2 + a^2}{R^2 + a^2} \tag{5}$$

where $R$ is the distance from the center of the galaxy, and we have assumed that the galactic center from the Sun, $R_0$, is 8.5kpc, the halo core radius, $a$, is 5kpc, and the local dark matter density, $\rho_0$, is $7.9 \times 10^{-3} \mathcal{M}_\odot/\mathrm{pc}^3$.

If we include all 6 stars, we find that red dwarfs make up $f = 0.44\%$ of the mass of the halo. We wish to examine the possibility that some of these stars are really disk stars contaminating the sample. To do this, we examine the luminosity distance, assuming a *high Z* disk metallicity of $[M/H] = -0.5$. [Note that if a star in the sample is a contaminating disk star, it would have high metallicity. (We thank C. Dahn for making this point to us)]. The four nearest stars then have heights above the galactic plane of $z = \{300, 800, 1500, 3400\}$pc. The nearest is clearly a disk star contaminating the sample and is thrown out. The next two are at a height above the plane where halo stars are about as common as disk stars, so we cannot determine their population. The fourth star is too far above the plane to be a disk star. Depending on whether or not we discard the two intermediate stars in addition to the closest star, the halo mass density drops to 0.14-0.37% of the halo, 16-40 times less than the reported result of BFGK.

[Note that inclusion of the sixth star with $(V - I) = 1.7$ (described above Eq. (3)) has changed the density very little, while extending the mass range to $0.2 \mathcal{M}_\odot$. Since this star is by far the brightest, its $d_{\max}$ is large, and its contribution to the mass fraction is very small, $\sim 0.03\%$].

*A Higher Metallicity Halo Calculation:* We have assumed that the stars seen in BFGK are extreme subdwarfs. Monet *et al.* (1992) also found a redder main sequence. We interpret this as a high metallicity population of disk stars. BFGK used the absolute magnitude of the reddest star in this population to derive their upper bound of 6%. In this section we can take the stars seen by BFGK to be in this redder sequence, yet take properties of each individual star (rather than assume they're all at the edge of the solar metallicity main sequence, as in BFGK). Stars with the same $(V - I)$ become brighter than in the low metallicity case; hence $d_{\max}$ becomes larger, increasing $M_{\text{halo},i}$ in equation (3). On the other hand, higher metallicity stars of a particular color are heavier, increasing $\mathcal{M}_i$ in equation (3). In addition, the apparent distance to the stars is greater so that more stars are included the sum. These opposing effects mean that the calculated halo density does not change strongly with assumed metallicity. For an assumed halo metallicity of $[M/H] = -0.5$, and including all six stars in the calculation, we find $f = 0.17\%$. If the observed stars are binary, then again, they are both more massive and brighter, so $f$ will only change slightly.

**Discussion and Conclusion:** Using the stellar data of extreme subdwarfs and models of low metallicity stars, we used several techniques to estimate the halo mass density of red dwarfs. First, from the data of BFGK, we obtained an upper bound by using properties of red dwarfs at the low metallicity main sequence edge: we found that these stars make up less than 0.95% of the halo with 95% confidence. Then we obtained an estimate of the contribution of red dwarfs to the halo from looking at the properties of the BFGK stars likely to be in the halo: we found that red dwarfs with $1.7 < (V - I) < 3$, i.e. mass $0.0875 < \mathcal{M} < 0.2\mathcal{M}_\odot$, make up 0.14-0.37% of the mass of the halo. If we use the latter estimate, assume a mass function of $\frac{dN}{d\mathcal{M}} \sim 1/\mathcal{M}$, and consider the entire mass range $0 < \mathcal{M} < 0.2\mathcal{M}_\odot$, we derive a halo mass fraction in pop. II lensing objects of only 0.25-0.67%. Thus, with these assumptions, the total mass of brown dwarfs and red dwarfs is roughly 0.25-0.67% of the mass of the halo, for a total mass of $(0.9 - 2.5) \times 10^9 \mathcal{M}_\odot$ out to 50 kpc assuming the halo model of equation (5).

This result disagrees with the MACHO report that a substantial fraction, $f = 18.8^{+33}_{-14}\%$ (95% confidence), of the halo is composed of c ompact halo objects (Alcock *et al.* 1995a). There are two ways to reconcile the two results: either the mass function must begin to rise more sharply than $\frac{dN}{d\mathcal{M}} \sim 1/\mathcal{M}$, or the halo model or mass function used by the MACHO group for their calculation must not be correct. It is possible for the mass function to rise sharply in the unsampled region below the hydrogen burning limit (or even have a spike at some particular mass). In addition, there may be a population of halo brown dwarfs formed under different conditions than the population II stars sampled here. There are strong degeneracies in any microlensing measurement; one cannot constrain the mass, velocity, impact parameter, and distance of a lens simultaneously. Thus, any microlensing mass fraction depends strongly on the halo model and mass function assumed (Alcock *et al.* 1995a,b; Kerrins 1995; Gates, Gyuk & Turner 1994). A no MACHO halo is not inconsistent with microlensing results since all lenses could be in a thick disk, or in the LMC (Alcock *et al.* 1995b; Sahu 1994a,b). Thus, although our result may cause us to question the "Standard halo model" result reported by the MACHO group, there are other possible interpretations.

We thank D. Richstone, G. Bernstein, M. Crone, C. Metzler, F. Adams, H. Harris, C. Stubbs, M. Turner, E. Gates, K. Greist and especially F. Allard and A. Gould for helpful discussions. We also thank C. Dahn for very useful comments as referee. We acknowledge support from the NSF Presidential Young Investigator Program, the NSF through grants

NSF PHY-9406745 and NSF PHY-9407194, and the Univ. of Michigan Physics Dept. K.F. thanks the Inst. for Theoretical Physics at Santa Barbara for hospitality.

**FIGURE CAPTIONS**

Figure 1. Figure reproduced from (Dahn *et al.* 1995) shows the H-R diagram of nearby stars with measured parallax. Filled circles are high metallicity disk stars, open circles are low metallicity halo subdwarfs. We have superimposed, as a solid line, the $[M/H] \equiv \log(Z/Z_\odot) = -1.5$ calculation of (Baraffe *et al.* 1995).

Figure 2. Points are the six reddest stars from BFGK. Masses were calculated by matching colors with masses using $[M/H] \equiv \log(Z/Z_\odot) = -1.5$ models of Baraffe *et al.* (1995). Distance moduli were calculated using the absolute magnitude-color relation of the extreme (low-metallicity) subdwarfs in Fig. 1 as fit by stellar modeling by Baraffe *et al.* (1995). The solid line is $d_{\max}$, the sensitivity of the observation by BFGK. There is a feature in $d_{\max}$ at $\mathcal{M} = 0.13\mathcal{M}_\odot$ because BFGK used different chips to observe different color ranges. Stars in the shaded region would be too dim to see.